\documentclass[aps,pre,twocolumn,groupedaddress]{revtex4-1}
\usepackage{bm}
\usepackage[]{natbib}
\usepackage{graphicx}
\usepackage{dcolumn}
\usepackage{amsmath,amsfonts,amsthm,amssymb}
\usepackage[caption=false]{subfig}
\usepackage{float}
\usepackage{tabularx}
\usepackage{hyperref}
\usepackage{enumerate}
\hypersetup{colorlinks=true,linkcolor=blue,citecolor=blue,urlcolor=blue}

\begin{document}


\title{Transitions between patterned states in vegetation models for semi-arid ecosystems}


\author{Karna Gowda}
\affiliation{Department of Engineering Sciences and Applied Mathematics, Northwestern University, Evanston, IL 60208, USA}

\author{Hermann Riecke}
\affiliation{Department of Engineering Sciences and Applied Mathematics, Northwestern University, Evanston, IL 60208, USA}

\author{Mary Silber}
\email[]{m-silber@northwestern.edu}
\affiliation{Department of Engineering Sciences and Applied Mathematics, Northwestern University, Evanston, IL 60208, USA}

\newcommand{\Det}{\text{Det}}
\providecommand{\abs}[1]{\lvert#1\rvert}
\providecommand{\norm}[1]{\lVert#1\rVert}

\date{\today}

\begin{abstract}
A feature common to many models of vegetation pattern formation in semi-arid ecosystems is a sequence of qualitatively different patterned states, ``gaps $\to$ labyrinth $\to$ spots'', that occurs as a parameter representing precipitation decreases. We explore the robustness of this ``standard'' sequence in the generic setting of a bifurcation problem on a hexagonal lattice, as well as in a particular reaction-diffusion model for vegetation pattern formation. Specifically, we consider a degeneracy of the bifurcation equations that creates a small bubble in parameter space in which stable small-amplitude patterned states may exist near two Turing bifurcations. Pattern transitions between these bifurcation points can then be analyzed in a weakly nonlinear framework. We find that a number of transition scenarios besides the standard sequence are generically possible, which calls into question the reliability of any particular pattern or sequence as a precursor to vegetation collapse. Additionally, we find that clues to the robustness of the standard sequence lie in the nonlinear details of a particular model.
\end{abstract}

\pacs{87.23.Cc, 89.75.Fb, 89.75.Kd, 05.45.-a}

\maketitle

\section{Introduction}
\begin{figure}[t]\centering
\includegraphics[width=\columnwidth]{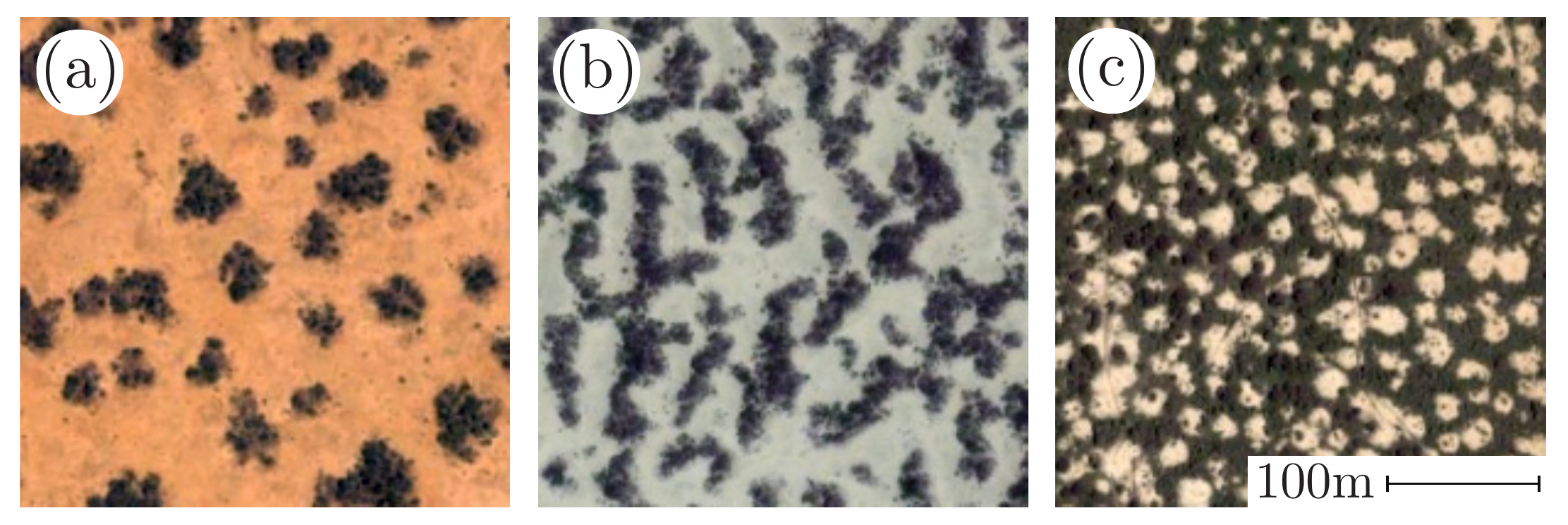}
\caption{(Color online) Aerial images of flat terrain vegetation patterns. (a) Spot patterns in Sudan (+11.582465, +27.913875), (b) tiger bush labyrinth patterns in Niger (+13.076354, +2.202780), (c) gap patterns in Senegal (+15.206133, -14.894417). Images $\copyright$ Google, DigitalGlobe.}
\label{fig:pattern_images}\end{figure}

Regular patterns in vegetation are among the most visually captivating and intriguing features of aerial images. These patterns appear ubiquitously in semi-arid ecosystems, where water is a limiting resource for vegetation growth. Examples include the regular stripe patterns that form on gradual slopes, and the spot, gap, and labyrinth patterns that form on flat terrain \cite{Borgogno:2009boa} (see Figure \ref{fig:pattern_images}). Flat terrain patterns are observed to be long-lived steady states and are thought to be self-organized phenomena that result from a symmetry-breaking instability in the underlying ecosystem dynamics \cite{Klausmeier:1999woa, Couteron:2001wt, HilleRisLambers:2001tia, vonHardenberg:2001bka}. This view has informed efforts over the last two decades to model vegetation pattern formation mathematically.

A number of models have been proposed to explain vegetation patterns in semi-arid ecosystems (see \cite{Borgogno:2009boa} for a review of work done prior to 2009). In particular, many deterministic PDE systems model vegetation patterns as self-organized phenomena that emerge due to positive feedbacks at short spatial ranges and competitive effects at long ranges \cite{Klausmeier:1999woa, vonHardenberg:2001bka, HilleRisLambers:2001tia, LeJeune:2004bma, Gilad:2004bp}. Short range feedbacks are due in part to nutrient and water related facilitation, while long range effects result from competition for limited resources \cite{Rietkerk:2008cda}. Typically, nonlinear terms in these models capture the local interactions between water and vegetation, while advective and diffusive terms describe spatial transport. Patterns emerge in many of these systems at a finite critical wavelength through a Turing instability \cite{Turing:1952vn}. In this paper, we focus on PDE models that exhibit this mechanism for pattern formation.

As the mean precipitation level of a system slowly diminishes, patterns may appear in vegetated ecosystems and then undergo a sequence of transitions between qualitatively different states before collapsing to a sparsely vegetated or barren state. Since a number of vegetation pattern models \cite{vonHardenberg:2001bka, Rietkerk:2002ufa, LeJeune:2004bma, Gilad:2004bp} feature a common ``standard'' sequence of patterns as precipitation decreases,
\begin{align}
		\text{gaps} \to \text{labyrinth} \to \text{spots},\label{eq:sequence}
\end{align}
it has been proposed that these characteristic patterns may serve as early indicators of a semi-arid ecosystem's imminent shift to desert \cite{Rietkerk:2004vqa, Scheffer:2009gga}. We will examine this proposition closely, first in the general setting of a degenerate equivariant bifurcation problem, and then in the model by von Hardenberg \textit{et al.} as an example. \cite{vonHardenberg:2001bka}.

Previous bifurcation studies by LeJeune \textit{et al.} \cite{LeJeune:2004bma}, Dijkstra \cite{Dijkstra:2011cx}, and Kealey \& Wollkind \cite{Kealy:2011isa} have analyzed 2D patterned states in vegetation models. One common approach likens spot, gap, and stripe patterns observed in nature to idealized patterns on a 2D hexagonal lattice. In particular, LeJeune \textit{et al.} \cite{LeJeune:2004bma} introduced a simple model featuring a patterned state between two Turing bifurcations. A bifurcation problem on a 2D hexagonal lattice was formulated and the coefficients of the amplitude evolution equations were computed in closed form. Considering a parameter set for which transitions between patterned states occurred entirely at small amplitude, an analog of the sequence \eqref{eq:sequence} was observed.

\begin{figure}\centering
\includegraphics[width=\columnwidth]{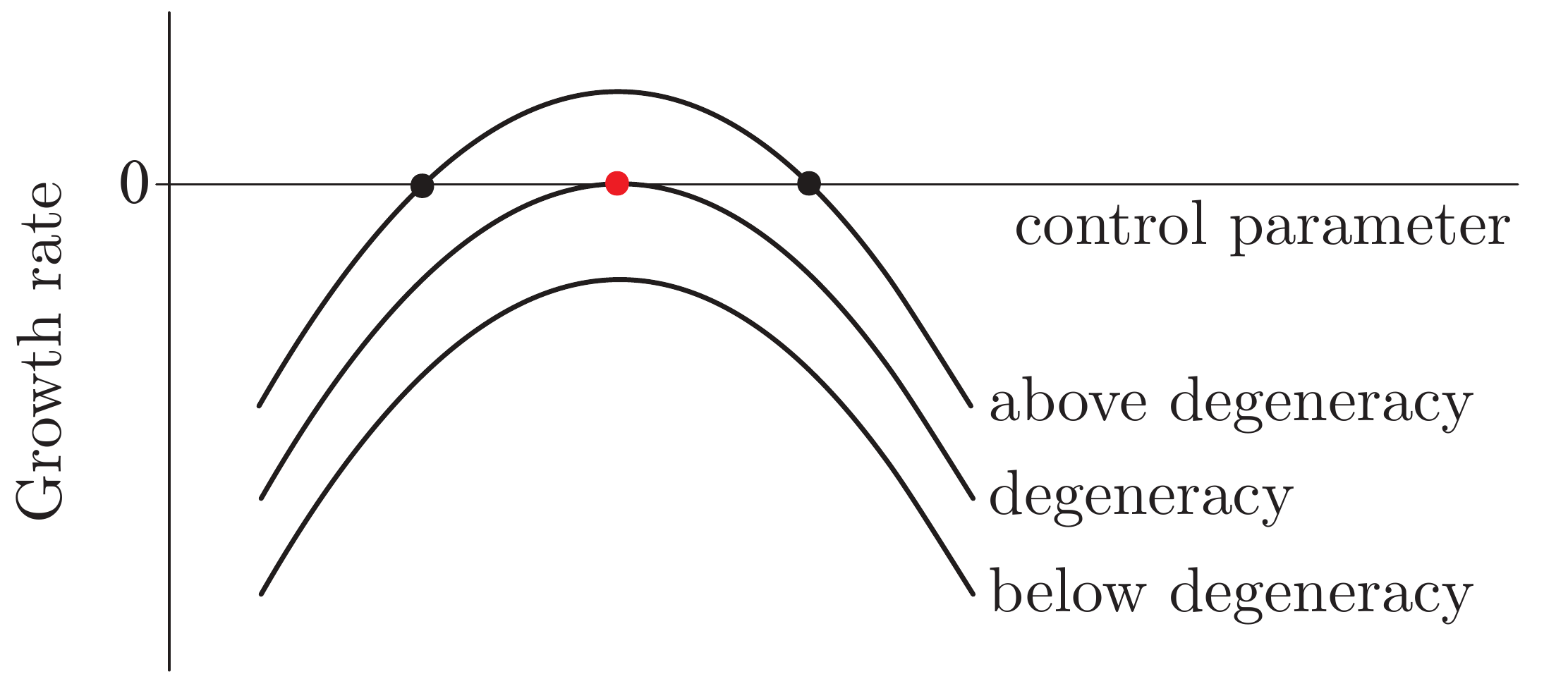}
\caption{Growth rate of the critical Fourier mode perturbing a spatially uniform equilibrium, plotted as a function of the control parameter. The points denote Turing bifurcations.}\label{fig:turing_degeneracy}
\end{figure}

Here, we introduce a framework for generic PDE systems in which pattern transitions between two Turing bifurcations can be analyzed at small amplitude. We formulate a finite-dimensional bifurcation problem on a 2D hexagonal lattice in the vicinity of a degenerate Turing bifurcation. This degeneracy is characterized by the growth rate of a critical Fourier mode perturbation to a spatially uniform state that varies with the control parameter as depicted in Figure \ref{fig:turing_degeneracy}. One can bring this degeneracy about by tuning a system so that two Turing bifurcations collapse to a single point in parameter space. In addition, we enforce a degeneracy that captures stable solutions to the amplitude equations. We then unfold these degeneracies so that two Turing bifurcations emerge at close proximity in parameter space and patterned states can be analyzed at small amplitude. By exploring the space of stable patterned solutions under this unfolding, we identify scenarios for transition between small-amplitude patterned states that may occur as a control parameter varies monotonically. We apply this framework to the model by von Hardenberg \textit{et al.} \cite{vonHardenberg:2001bka}, treating precipitation as a control parameter and performing weakly nonlinear analyses and numerical simulations near a degenerate Turing point.

This paper extends the analysis by LeJeune \textit{et al.} \cite{LeJeune:2004bma} to a general framework for PDE systems with patterned states arising from Turing instabilities, which allows us to explore pattern transition scenarios outside of the constraints imposed by a particular model. Using this framework, we find that an analog of sequence \eqref{eq:sequence} is only one of many possible scenarios. Alternative scenarios include transitions that exclude labyrinth patterns altogether, as well as scenarios involving only spot or only gap patterns. The existence of these alternatives calls into question the utility of sequence \eqref{eq:sequence} as a leading indicator of transition to desert, as any of these other sequences may arise depending on system details.

Additionally, this framework offers a method for identifying which characteristics of particular models are robust, and which are specific to model parameters. By computing amplitude equation coefficients as functions of the model parameters, one can make general statements about the restrictions a model imposes on pattern transitions that occur as a control parameter varies. In the case of the von Hardenberg \textit{et al.} model, the amplitude equation coefficients reveal that no small-amplitude solutions are stable near Turing points for the parameter set considered in \cite{vonHardenberg:2001bka, Meron:2004dra}, but that a small-amplitude analog of \eqref{eq:sequence} occurs for other distinguished parameter sets. Interestingly, we find that for other parameter sets this model can also exhibit a scenario involving only spot patterns, but that this behavior appears restricted to a small interval of the parameter we vary. Overall, our analysis offers some evidence that sequence \eqref{eq:sequence} is relatively robust in the model by von Hardenberg \textit{et al.} \cite{vonHardenberg:2001bka}.

This paper is structured as follows. Section \ref{sec:motivation} reviews the model by von Hardenberg \textit{et al.} \cite{vonHardenberg:2001bka} to motivate our analysis. Section \ref{sec:bifurcation} summarizes the formulation of a bifurcation problem on a hexagonal lattice and presents the degeneracies that form the basis of our analysis. Section \ref{sec:degeneracy} explores the parameter space of our perturbed degenerate bifurcation problem to identify transition scenarios in the interesting cases where solutions are stable at small amplitude. Section \ref{sec:model_degeneracy} returns to the model by von Hardenberg \textit{et al.} \cite{vonHardenberg:2001bka} to identify pattern transitions near Turing bifurcation points. The results of numerical simulations are also presented. Section \ref{sec:discussion} then concludes with a discussion of our results in the context of the current and future work.

\section{Motivating Example}\label{sec:motivation}
\begin{figure}\centering
\includegraphics[width=\columnwidth]{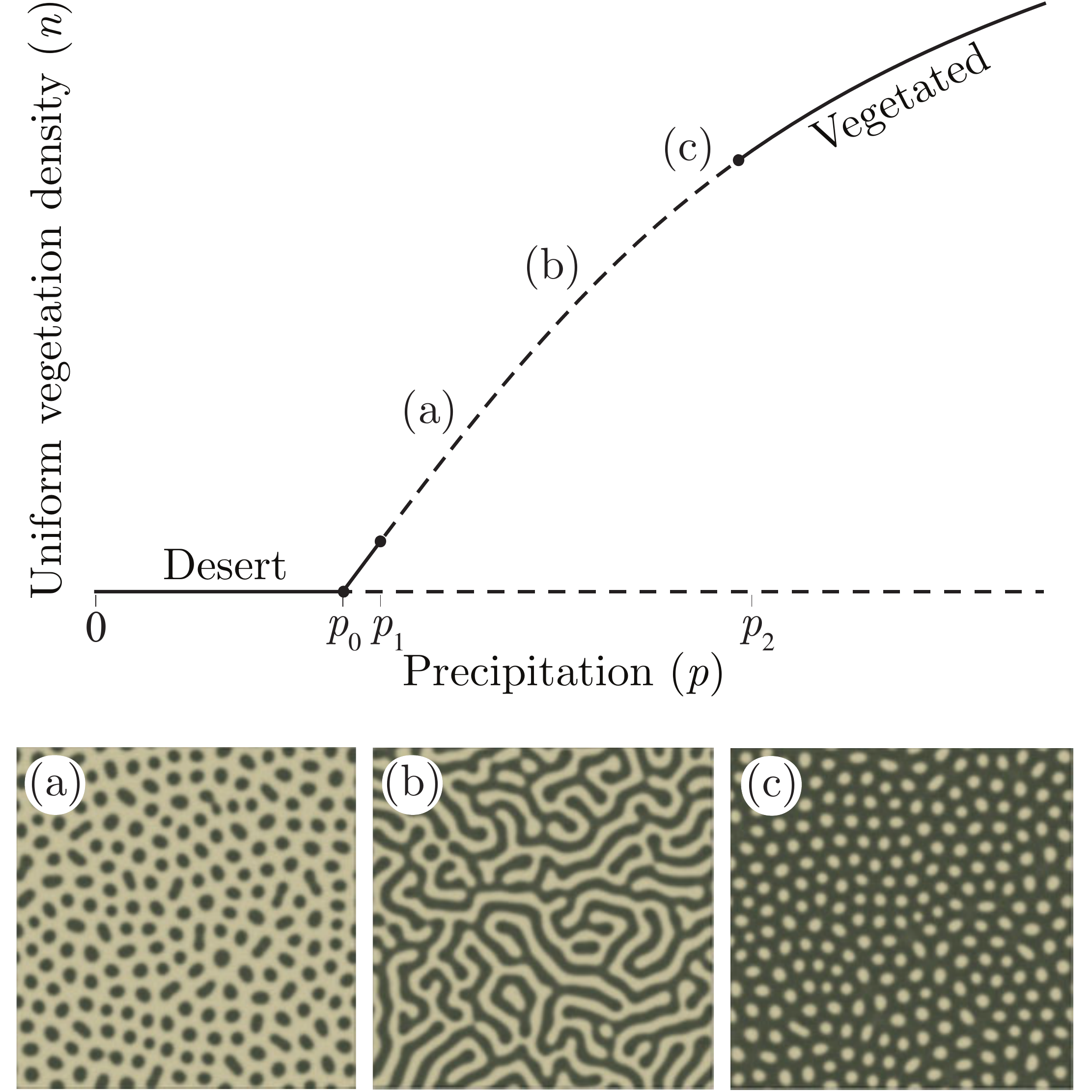}\caption{(Color online) Bifurcation diagram depicting the spatially uniform states of \eqref{eq:vonH} with numerical solutions, using the default parameter set $\gamma = \sigma = 1.6$, $\nu = 0.2$, $\rho=1.5$, $\beta = 3$ and $\delta=100$ \cite{vonHardenberg:2001bka, Meron:2004dra}. The desert state loses stability to a vegetated state at $p=p_0\approx 0.157$. The vegetated equilibrium is unstable to perturbations $n,w \sim e^{iqx}$ in the dashed region $p_1 < p < p_2$, where $p_1 \approx 0.169$ and $p_2 \approx 0.413$ are the Turing points. Numerical simulations use precipitation values (a) $p=0.20$, (b) $p=0.30$, and (c) $p=0.40$, with higher vegetation density plotted in darker shading.}
\label{fig:vonH_uniform}\end{figure}

To motivate the framework for our subsequent analysis, we briefly review the model by von Hardenberg \textit{et al.} \cite{vonHardenberg:2001bka} as an example. A detailed description of model terms, scaling, and behavior is given in \cite{Meron:2004dra}. In nondimensional form, the model is:
\begin{equation}
	\begin{aligned}
		n_t &= f(n,w) + \nabla^2 n,\\
		w_t &= g(n,w) + \delta \nabla^2(w - \beta n),\label{eq:vonH}
	\end{aligned}
\end{equation}
with 
\begin{equation}
	\begin{aligned}
		f(n,w) &= \left(\frac{\gamma w}{1 + \sigma w} - \nu \right)n - n^2,\\
		g(n,w) &= p - (1 - \rho n)w - w^2 n.\label{eq:vonH_nlin}
	\end{aligned}
\end{equation}
The variable $n(x,y,t)$ represents vegetation density and $w(x,y,t)$ represents ground water density. The nonlinear functions \eqref{eq:vonH_nlin} capture the effects of facilitation and competition. Spatial terms model the diffusive spread of vegetation and the transport of water, with cross-diffusion accounting for suction by plant roots in the latter. A parameter representing precipitation, $p$, imposes resource scarcity. Additionally, $\gamma$, $\sigma$, $\nu$, $\rho$ and $\beta$ are $\mathcal{O}(1)$ positive parameters and $\delta$, which characterizes the diffusivity of water relative to that of vegetation, is usually taken to be a large, positive parameter. One set of default parameter values examined in \cite{vonHardenberg:2001bka, Meron:2004dra} is given in the caption of Figure \ref{fig:vonH_uniform}.

\begin{figure}\centering
\includegraphics[width=\columnwidth]{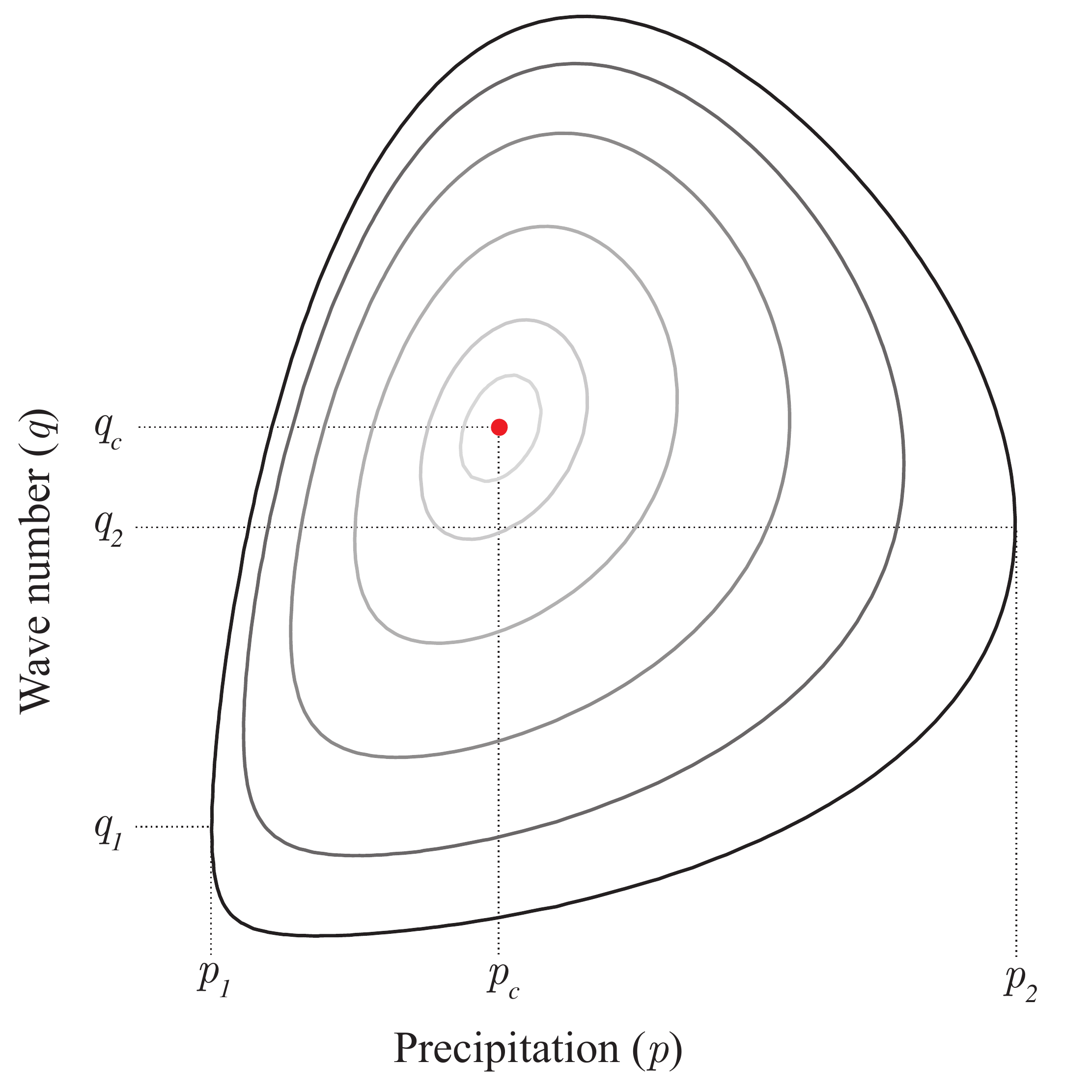}
\caption{Linear stability boundary of uniform vegetated states of \eqref{eq:vonH} plotted in the $p$-$q$ plane, where $q$ is the perturbing wave number. The boundary forms a closed bubble whose left and right endpoints are the Turing points $p_1$ and $p_2$, which have corresponding critical wave numbers $q_1$ and $q_2$. As the diffusion coefficient $\delta$ is decreased, the stability bubble collapses to a degenerate Turing point, $(p_c,q_c)$.}
\label{fig:vonH_bubble}\end{figure}

Spatially uniform equilibria $(n^*,w^*)$ of \eqref{eq:vonH} solve $f(n^*,w^*) = g(n^*,w^*) = 0$. One such equilibrium, corresponding to a non-vegetated ``desert'' state, is given by $(n^*,w^*)=(0,p)$. The desert state is linearly stable for $p<p_0$, and undergoes a transcritical bifurcation at $p=p_0$ to a uniform ``vegetated'' equilibrium, for which $n^*>0$. The bifurcation diagram depicting these spatially uniform states and their stability as a function of $p$ is plotted in Figure \ref{fig:vonH_uniform}. Turing bifurcations on the uniform vegetated equilibrium produce patterned states at the points $p=p_1,p_2$. One can determine these points by linearizing \eqref{eq:vonH} about the uniform vegetated equilibrium and examining the growth rate of Fourier mode perturbations $(n-n^*),(w-w^*) \sim e^{iqx}$, for arbitrary perturbing wave number $q$. From this, one obtains a stability boundary for the spatially uniform state in the $p$-$q$ plane, which in this case forms a closed bubble that is plotted in Figure \ref{fig:vonH_bubble}. The Turing points $p_1$ and $p_2$ are the left and right endpoints of this stability bubble, and they are each associated with a preferred critical wave number $q_1,q_2>0$.

Numerical simulations of \eqref{eq:vonH} yield asymptotic states that follow sequence \eqref{eq:sequence} for decreasing values of $p$ (see Figure \ref{fig:vonH_uniform}). It is natural to liken these solutions to regular patterns on a 2D hexagonal lattice \cite{LeJeune:2004bma, Kealy:2011isa}. A uniform vegetated state that develops dry gaps resembles a ``down-hexagons'' pattern ($H^-$), a labyrinthine intermediate state resembles a distorted stripes pattern ($S$), and a state of isolated vegetation spots resembles ``up-hexagons'' ($H^+$). Aspects of pattern existence and stability can be assessed at small amplitude in the weakly nonlinear regime surrounding a Turing bifurcation point, which we explore further for this model in Section \ref{sec:model_degeneracy}.

\section{Formulating the bifurcation problem}\label{sec:bifurcation}
Near a Turing point $(p_{crit},q_{crit})$, the behavior of a solution to a pattern-forming system is characterized by the modes of a Fourier expansion corresponding to wave number $q_{crit}$. Aspects of pattern formation can be analyzed through the time-varying amplitudes of these critical modes. The form of the amplitude equations for Fourier modes on a 2D hexagonal lattice can be derived through a standard calculation, described for instance in \cite{Golubitsky:2012ud, Cross:1993tka, Hoyle:2006ur}. Here, we summarize key points and results of this derivation, and then formulate the degeneracies that allow us to analyze transitions between patterned states at small amplitude.

We consider the evolution of critical Fourier modes on a hexagonal lattice perturbing the uniform state,
\begin{align}
		z_1 e^{i\mathbf{q_1}\cdot \mathbf{x}} + z_2 e^{i\mathbf{q_2}\cdot \mathbf{x}} + z_3 e^{i\mathbf{q_3}\cdot \mathbf{x}} + c.c. + \cdots.\label{eq:crit_mode_amps}
\end{align}
The wave vectors $\mathbf{q_1},$ $\mathbf{q_2},$ $\mathbf{q_3}$ are chosen such that
\begin{equation}
	\begin{aligned}
		\mathbf{q_1} &= q_{crit}(1,0),\\
		\mathbf{q_2} &= q_{crit}\left(-1/2, \sqrt{3}/2 \right),\\
		\mathbf{q_3} &= -(\mathbf{q_1} + \mathbf{q_2}).
	\end{aligned}
\end{equation}
These vectors lie on the critical circle $\abs{\mathbf{q}} = q_{crit}$ and $z_1$, $z_2$, $z_3$ (as well as their complex conjugates) are the complex amplitudes of the corresponding critical modes. Near $p_{crit}$, all other complex Fourier modes associated with the hexagonal lattice are linearly damped.

The form of the equations describing the evolution of the critical mode amplitudes near a Turing point can be determined using an equivariant bifurcation theory approach \cite{Golubitsky:2012ud}. To cubic order, these equations are
\begin{equation}
	\begin{aligned}
	\dot{z}_1 = \mu z_1 &+ a \bar{z}_2 \bar{z}_3\\
	 &- \left(b \vert z_1 \vert^2 + c (\vert z_2 \vert^2 + \vert z_3 \vert^2)\right)z_1,\\
 	\dot{z}_2 = \mu z_2 &+ a \bar{z}_1 \bar{z}_3\\
 	 &- \left(b \vert z_2 \vert^2 + c (\vert z_1 \vert^2 + \vert z_3 \vert^2)\right)z_2,\\
 	\dot{z}_3 = \mu z_3 &+ a \bar{z}_1 \bar{z}_2\\
 	 &- \left(b \vert z_3 \vert^2 + c (\vert z_1 \vert^2 + \vert z_2 \vert^2)\right)z_3.
	 \label{eq:bifeqns1}
	\end{aligned}
\end{equation}
The coefficients $\mu, a, b,$ and $c$ are real-valued and are determined by the specific terms and parameter values of a system in a neighborhood of a Turing point. These equations describe the branching and relative linear stability of stripes and hexagons solutions. Specifically, the relative stability of a solution to lattice perturbations is governed by the eigenvalues of the linearization of \eqref{eq:bifeqns1} about that solution. Solutions and their eigenvalues are listed in Table \ref{tab:hex_solns}. From these eigenvalues, one may determine conditions for the stability of a solution. For example, a necessary condition for the stability of stripes is $b>0$, which ensures that the eigenvalue $-2 b x_s^2$ is negative. Eigenvalues are often repeated (indicated by $\times 2$ in Table \ref{tab:hex_solns}) due to the spatial symmetries of a solution, and zero eigenvalues reflect the neutral stability of solutions to translation.

\begin{table}
	\caption{Branching equations for solutions on a hexagonal lattice, and eigenvalues of linearizations of \eqref{eq:bifeqns1}, together with their multiplicities. $\mathbf{z} = (z_1,z_2,z_3)$, and $x_s, x_h, x_1,x_2 > 0$.}\label{tab:hex_solns}
    \begin{tabular}{ll}
    \hline
    Solution and branching equation     & Eigenvalues	            \\ \hline
    Stripes ($S$)                       & $-2bx_s^2
	$, $0$,                \\
    $\mathbf{z} = (x_s, 0, 0)$          & $(b-c)x_s^2+a x_s$ ($\times 2$),\\
    $0 = \mu x_s - b x_s^3$             & $(b-c)x_s^2-a x_s$ ($\times 2$)\\
                                        &                               \\
    Up-hexagons ($H^+$)                 & $-3a x_h$, 0 ($\times 2$),	    \\
    $\mathbf{z} = (x_h, x_h, x_h)$      & $a x_h - 2(b+2c)x_h^2$, 		\\
    $0=\mu x_h + a x_h^2 - (b+2c)x_h^3$ & $-2a x_h - 2(b-c)x_h^2$($\times 2$)\\
                                        &                               \\
    Down-hexagons ($H^-$)               & $3a x_h$, 0 ($\times 2$),      \\
    $\mathbf{z} = -(x_h, x_h, x_h)$     & $-a x_h - 2(b+2c)x_h^2$,      \\
    $0=\mu x_h - a x_h^2 - (b+2c)x_h^3$ & $2a x_h - 2(b-c)x_h^2$ ($\times 2$)\\
                                        &                               \\
    Mixed-modes ($MM$)                  & Always unstable               \\
    $\mathbf{z} = (x_1, x_2, x_2)$      &                               \\
    $0=\mu x_1 + a x_2^2 - (b x_1^2 + 2 c x_2^2)x_1$ &                  \\
	$0=\mu x_2 + a x_1 x_2 - (b+c)x_2^3 - c x_1^3$ &					\\ \hline
    \end{tabular}
\end{table}

We now introduce a distinguished control parameter $\lambda$. The coefficients in \eqref{eq:bifeqns1} are generally functions of this parameter, i.e. $\mu = \mu(\lambda)$, etc., which we constrain to vary with $\lambda$ in the following way. First, we force the bifurcation problem to occur in the vicinity of a degenerate Turing point at $\lambda = 0$ (without loss of generality), which corresponds to $\mu$ varying quadratically in $\lambda$ to leading order (i.e. $\mu(\lambda) = \mu''(0) \lambda^2/2 + \cdots$, where $\mu''(0) < 0$). Unfolding the degenerate point results in the expansion
\begin{align}
	\mu(\lambda) = \mu_0 + \mu_1 \lambda + \mu_2 \lambda^2 + \cdots,
\end{align}
where $\mu_0$ and $\mu_1$ are small parameters and $\mu_2 < 0$. This unfolding corresponds to a small parameter perturbation to a Turing degeneracy, and two Turing points may emerge as real solutions to $\mu(\lambda) = 0$. An illustration of the Turing instability growth rate, which is proportional to $\mu$, is shown near a Turing degeneracy in Figure \ref{fig:turing_degeneracy}. Second, if the quadratic coefficient $a \neq 0$, it can be shown that all solutions to \eqref{eq:bifeqns1} bifurcate unstably \cite{Golubitsky:2012ud}. Hence, to capture stable solutions of \eqref{eq:bifeqns1}, it is standard to consider \eqref{eq:bifeqns1} near $a = 0$. We therefore unfold $a$ as
\begin{align}
	a(\lambda) = a_0 + a_1 \lambda + \cdots,
\end{align}
where $a_0$ is another small parameter. Last, we do not consider the variation of $b$ or $c$ with $\lambda$, which is reasonable if we avoid degeneracies involving those coefficients.

Through a rescaling of time and amplitudes and ignoring higher order terms, we can write unfoldings of the degenerate linear and quadratic coefficients of \eqref{eq:bifeqns1} as
\begin{align}
		\mu(\lambda) = \mu_0 + \mu_1 \lambda -\lambda^2,\, a(\lambda) = a_0 + \text{sgn}(a_1) \lambda,\label{eq:unfolded_coeffs}
\end{align}
where $\mu_0$, $\mu_1$ and $a_0$ are small parameters. The equations \eqref{eq:bifeqns1} with coefficients \eqref{eq:unfolded_coeffs} now show transitions between pattern solutions that can occur entirely at small amplitude.

\section{Pattern transitions near Turing degeneracy}\label{sec:degeneracy}
For fixed values of the coefficients $b$ and $c$, the eigenvalues in Table \ref{tab:hex_solns} specify regions in the $a$-$\mu$ plane where solutions to \eqref{eq:bifeqns1} are stable, and the coefficients \eqref{eq:unfolded_coeffs} describe paths through these regions that are parameterized by the control parameter $\lambda$. A path through the stability region of a solution corresponds to the existence of a stable solution for an interval of the control parameter. Exiting a region through a stability boundary represents a solution losing stability and points to a transition between patterned states. In the following, we outline our procedure for identifying small-amplitude pattern transition scenarios.

By requiring the non-identically zero eigenvalues in Table \ref{tab:hex_solns} to be negative, stability regions in the $a$-$\mu$ plane for solutions to \eqref{eq:bifeqns1} divide consideration into three distinct cases: (i) $0<b<c$, (ii) $-c<b<0$, and (iii) $-b/2 < c< b$. Together, these make up all cases in which stable solutions to \eqref{eq:bifeqns1} exist. Notably, case (i) is the only case in which small-amplitude stripes are stable.

\begin{figure}\centering
\includegraphics[width=\columnwidth]{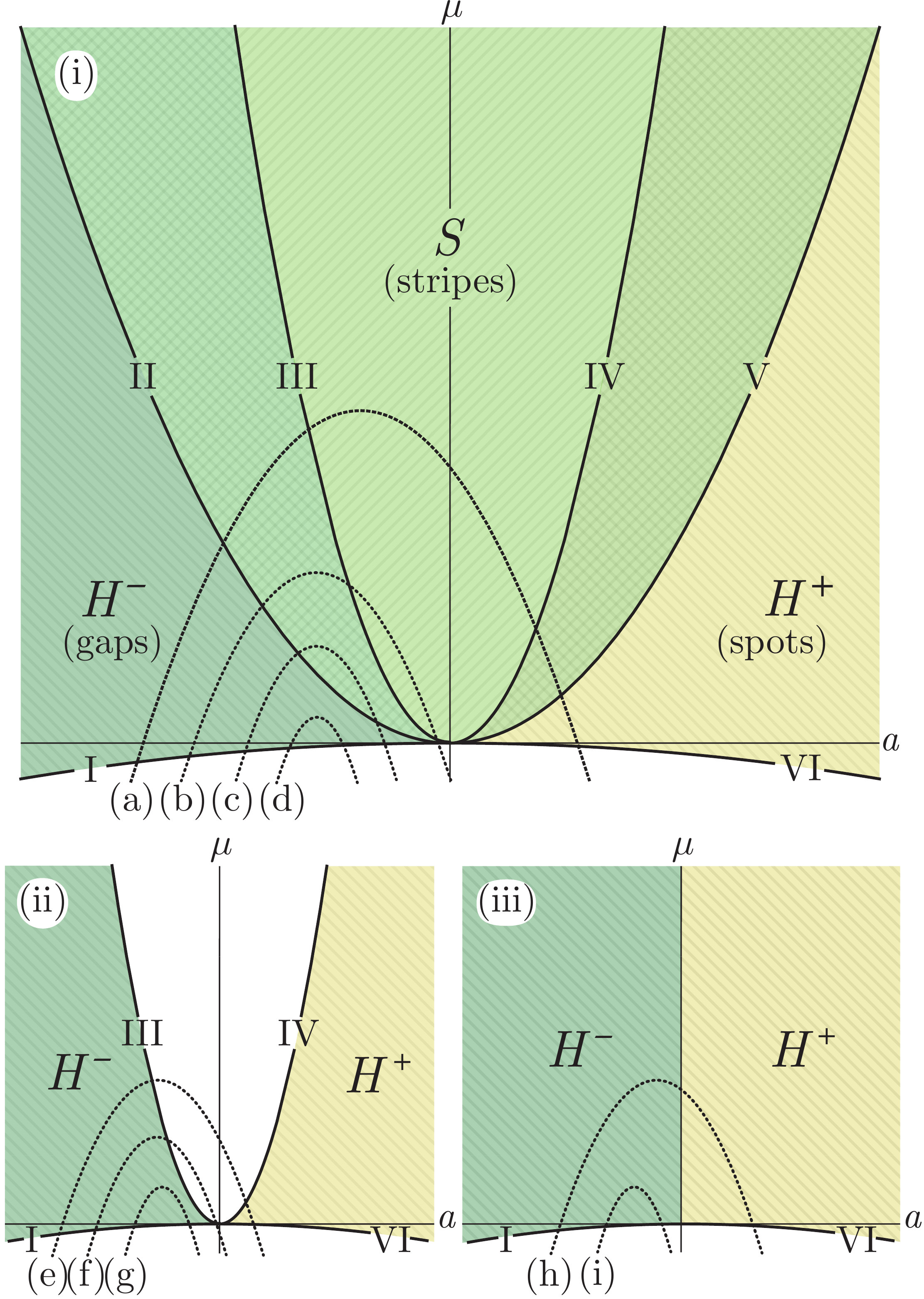}
\caption{(Color online) Stability regions in the $a$-$\mu$ plane for $S$ and $H^{+/-}$ solutions to \eqref{eq:bifeqns1} for cases (i) $0<b<c$, (ii) $-c<b<0$ and (iii) $-b/2<c< b$, with sample paths (a-i) formed by varying $\lambda$ in \eqref{eq:unfolded_coeffs} (discussed in the text). The boundaries $H_1$, $H_2$, and $S_1$ given in \eqref{eq:region_bdrys} are labeled I-VI and distinguish between boundaries on either side of $a=0$. Cross hatching denotes regions of bistability between $S$ and $H^{+/-}$ solutions. Region colors reflect a vegetation health gradient, where dark green reflects gap vegetation and yellow reflects less healthy spot vegetation.}
\label{fig:regions}\end{figure}

Table \ref{tab:regions} lists stability regions for each case, which are bounded by the curves
\begin{equation}
	\begin{aligned}
		\text{$H_1$: } &\mu = -a^2/4(b+2c),\\
		\text{$H_2$: } &\mu = a^2(2b+c)/(b-c)^2,\\
		\text{$S_1$: } &\mu = a^2b/(b-c)^2.
	\end{aligned}\label{eq:region_bdrys}
\end{equation}
$H_1$ is derived from the existence condition for hexagons, and $H_2$ and $S_1$ come from necessary conditions for stability of hexagons and stripes, respectively. These stability regions are plotted in Figure \ref{fig:regions}, and the bounding curves are labeled I-VI and distinguish between regions of stability on either side of the line $a=0$.

\begin{table}
	\caption{Regions of stability for stripes and hexagon solutions to \eqref{eq:bifeqns1} in the $a$-$\mu$ plane for cases (i) $0<b<c$, (ii) $-c<b<0$, and (iii) $-b/2<c<b$. Regions are bounded by the curves given in \eqref{eq:region_bdrys}. $H^+$ is stable when $a>0$, and $H^-$ is stable when $a<0$.}\label{tab:regions}
    \begin{tabularx}{\columnwidth}{X c c c}
    \hline
    				& Case (i) 			& Case (ii) 	& Case (iii)\\ \hline
    Stripes ($S$)  	& $\mu > S_1$ 	& --- 	& ---\\
    Hexagons ($H^{+/-}$)& $H_1<\mu<H_2$ &  $H_1< \mu <H_2$ & $\mu>H_1$\\
	\hline
    \end{tabularx}
\end{table}

\begin{table}
	\caption{Distinct boundary crossing sequences and inferred transition scenarios for cases (i) $0<b<c$, (ii) $-c<b<0$,  and (iii) $-b/2<c<b$, with example paths from Figure \ref{fig:regions} indicated. Dashes in case (ii) indicate that no intermediate state is stable at small amplitude. ``Symmetric'' sequences occuring for $a>0$ (e.g. VI,VI) are omitted.}\label{tab:sequences1}
	\begin{tabularx}{\columnwidth}{X l X l}\hline
	    & Ex. & Sequence     	& Transition scenario 	\\ \hline
	    Case (i):&      & 		&	\\
	    & (a) &I-VI          	& $H^- \to S \to H^+$,	\\
	    & (b) &I,II,III,III,II,I            & $H^- \to S \to H^-$	\\
	    & (c) &I,II,II,I      & $H^-$ (w/ $S$ bistability)	\\
	    & (d) &I,I   & $H^-$	\\
	
	    Case (ii):&             & 	&								\\
	    & (e) &I,III,IV,VI    & $H^- \to - \to H^+$,				\\
	    & (f) &I,III,III,I            & $H^- \to - \to H^-$					\\
	    & (g) &I,I    & $H^-$				\\
	
	    Case (iii):&        &              & 						\\
	    & (h) &I,VI         	& $H^- \to H^+$,					\\
	    & (i) &I,I            & $H^-$								\\\hline
    \end{tabularx}
\end{table}

\begin{figure}\centering
\includegraphics[width=\columnwidth]{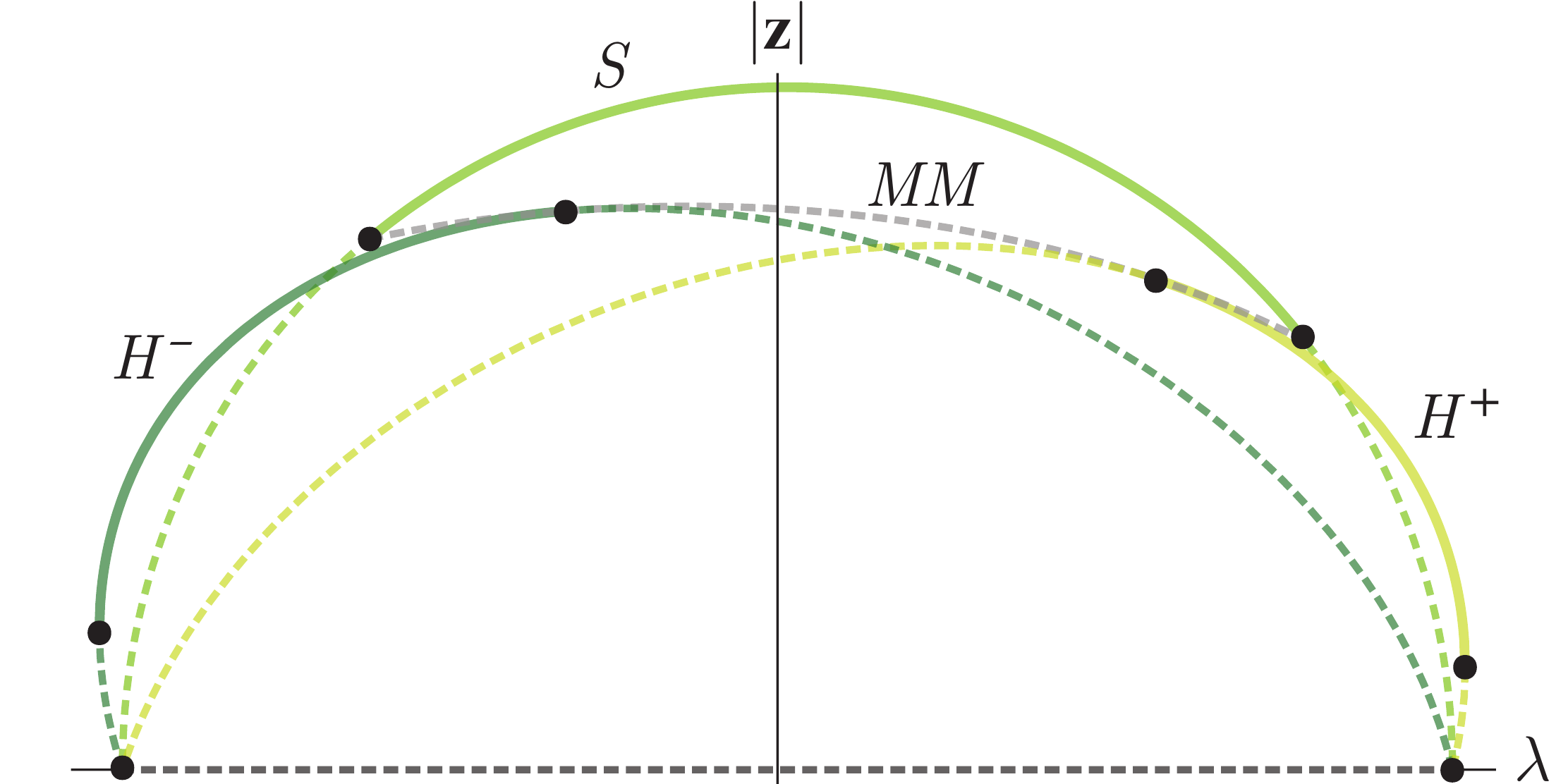}
\caption{(Color online) Schematic bifurcation diagram corresponding to path (a) in Figure \ref{fig:regions}, taking $\text{sgn}(a_1)=1$. The Euclidean norm of $\mathbf{z} = (z_1, z_2, z_3)$ is plotted, with stable solutions solid and unstable solutions dashed. Bifurcation points are indicated by dots.}\label{fig:example}
\end{figure}
 
Varying the control parameter $\lambda$ in the coefficients \eqref{eq:unfolded_coeffs} leads to parabolic paths in the $a$-$\mu$ plane. Depending on the sign of $a_1$ and the small parameters $\mu_0$, $\mu_1$, $a_0$, these parabolic paths cut through the regions of pattern existence/stability in Table \ref{tab:regions}. Some example paths through these regions are sketched in Figure \ref{fig:regions}. We observe that example path (a) crosses through boundaries I-VI. This means varying $\lambda$ along this path leads \eqref{eq:bifeqns1} sequentially through the regions of stability for $H^-$, $S$, and $H^+$ solutions. A bifurcation diagram corresponding to this path is sketched in Figure \ref{fig:example}, from which we infer the transition sequence ``$H^- \to S \to H^+$'' (i.e. ``gaps $\to$ labyrinth $\to$ spots'') occurs as $\lambda$ varies.

It follows from this example that for each case, each distinct boundary crossing sequence is linked to a pattern transition sequence. Distinct boundary crossings and inferred transitions for all three cases are listed in Table \ref{tab:sequences1}. ``Symmetric'' sequences occur where boundaries are crossed on one side of the line $a=0$ only (e.g. case (i) sequence ``I,I''), while ``asymmetric'' sequences occur where boundaries are crossed on both sides of the $\mu$-axis (e.g. case (i) sequence ``I,II,III,IV,V,VI''). Path (a) in Figure \ref{fig:regions} results in an asymmetric sequence, while paths (b-d) result in symmetric sequences. Only one asymmetric sequence, corresponding to a transition involving both $H^+$ and $H^-$ solutions, exists for each case.

One type of scenario not explicitly described above occurs when a parabolic path crosses through a region of hexagon stability without crossing the line $\mu = 0$. Since Turing bifurcations occur when $\mu = 0$, such a scenario can be interpreted as a small-amplitude hexagon states that occurs without Turing bifurcations. These states coexist with spatially uniform states that are stable to Fourier mode perturbations, and thus may only be accessed through a finite amplitude perturbation. Transition scenarios for such patterned states must necessarily involve only one type of hexagon solution.

This analysis shows that a number of transition scenarios between patterned states, beyond the standard ``gaps $\to$ labyrinth $\to$ spots,'' are possible. Many of these scenarios correspond to ``symmetric'' sequences and therefore only involve one type of hexagons solution (e.g. case (i) sequence ``I,II,III,III,II,I''). This demonstrates that it is possible for pattern transitions to occur without the appearance of spots, for instance, as in paths (b-d) in Figure \ref{fig:regions}. The occurence of spots as precursors to pattern collapse may therefore be sensitive to the specific parameters of a system.

\section{Pattern transitions in the model by von Hardenberg \textit{et al.}}\label{sec:model_degeneracy}

In this section, we analyze patterned states in the model \eqref{eq:vonH} by von Hardenberg \textit{et al.} \cite{vonHardenberg:2001bka}. Note that we only consider the dependence of pattern transition scenarios on the diffusion parameter $\delta$ as an illustration, but that one could just as well consider the dependence on other parameters. We obtain the coefficients of the amplitude equations \eqref{eq:bifeqns1} following the results of Judd and Silber \cite{Judd:2000wma}, who derive these coefficients perturbatively for general two-component reaction-diffusion systems with diagonal diffusion. \eqref{eq:vonH} is transformed into this problem by diagonalizing the diffusion matrix. 

For the default parameter set $\gamma = \sigma = 1.6$, $\nu = 0.2$, $\rho=1.5$, $\beta = 3$ and $\delta=100$, Turing instabilities occur along the spatially uniform vegetated equilibrium at $(p_1, q_1) \approx (0.169, 0.106)$ and $(p_2, q_2) \approx (0.414, 0.206)$. The coefficients $b$ and $c$ of \eqref{eq:bifeqns1} are negative at both Turing points, which renders unstable small-amplitude stripes and hexagons. Weakly nonlinear theory is therefore unable to describe stable patterned solutions to \eqref{eq:vonH} near these points. Numerical simulations, which are shown in Figure \ref{fig:vonH_uniform}, reveal that the ``gaps $\to$ labyrinth $\to$ spots'' sequence occurs at large amplitude as precipitation decreases.

\begin{figure}\centering
\includegraphics[width=\columnwidth]{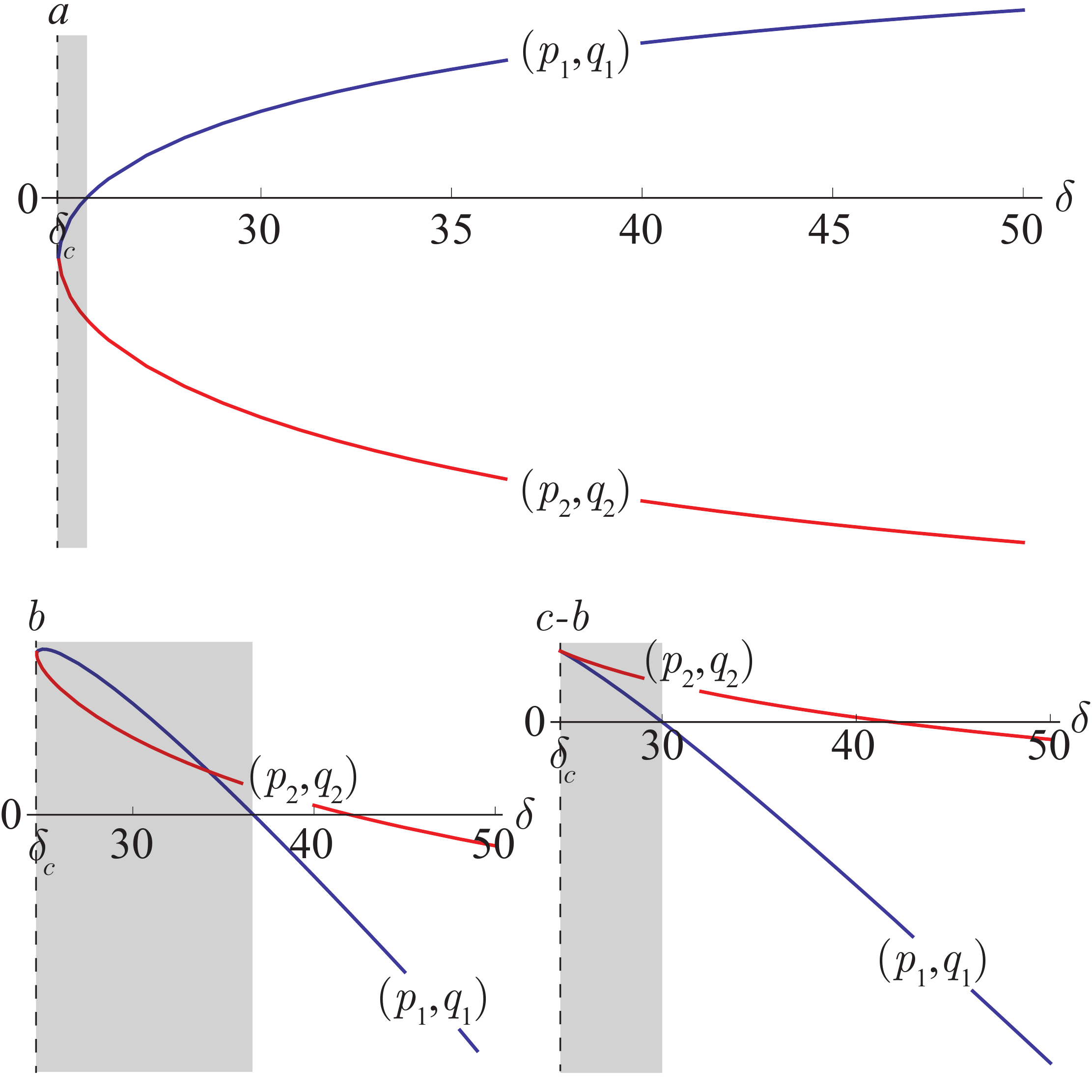}
\caption{(Color online) The coefficients $a$, $b$, and $c-b$, plotted at Turing points $(p_1, q_1)$ and $(p_2, q_2)$ as functions of the parameter $\delta$ (default parameter values used otherwise). Grey-shaded intervals indicate where $a<0$, $b>0$, and $c-b>0$ at both Turing points simultaneously.}
\label{fig:vonH_coeffs}\end{figure}

\begin{figure}\centering
\includegraphics[width=\columnwidth]{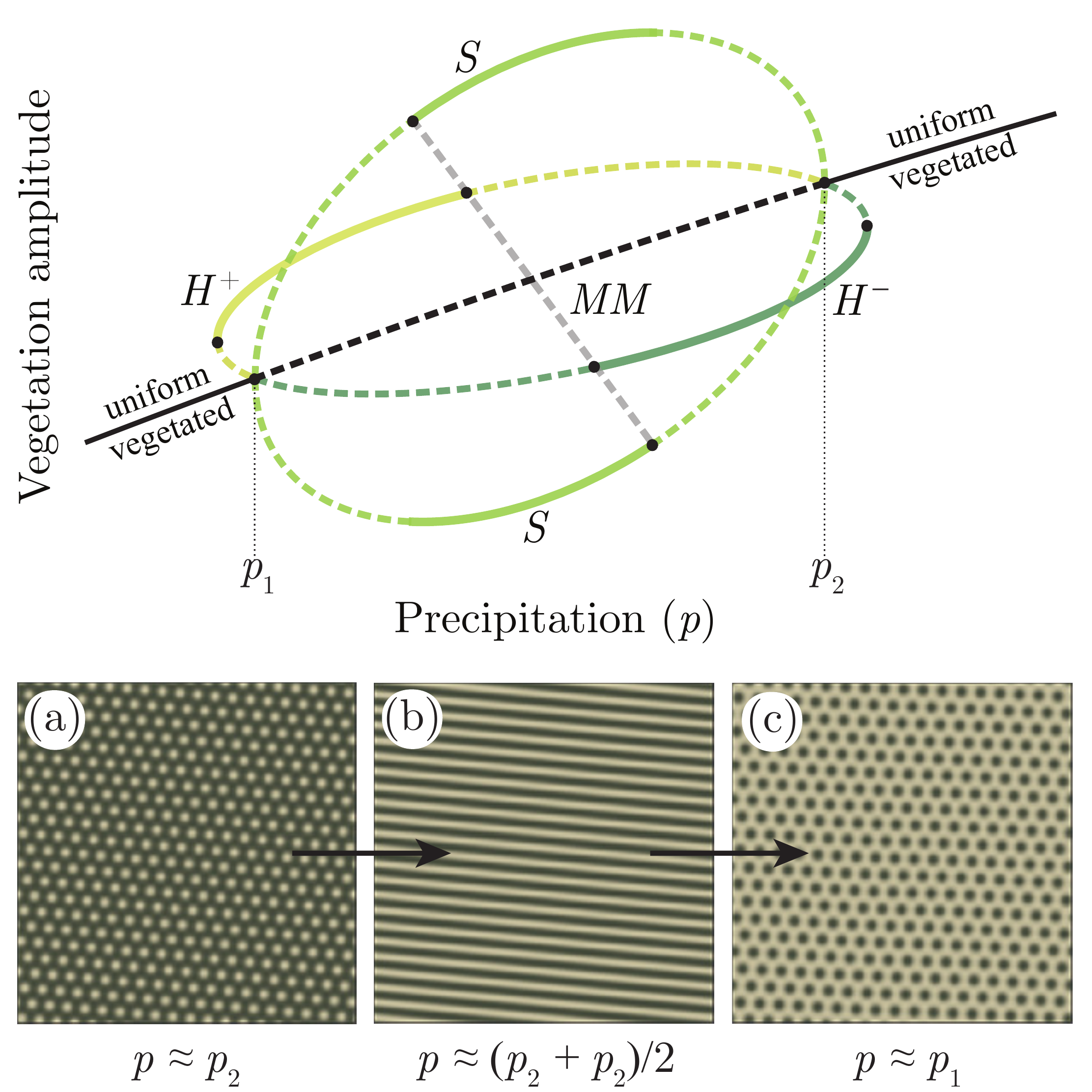}
\caption{(Color online) Schematic bifurcation diagram for solutions of \eqref{eq:vonH} with $\delta \in [25.5,30.0]$ and numerical solutions with $\delta = 27.0$ (default parameter values used otherwise). The numerical simulation (a) is evolved to a steady state resembling an $H^-$ solution from spatially noisy initial conditions at $p = 0.292 \approx p_2$. $p$ is then decreased to $0.259 \approx (p_1 + p_2)/2$ and the solution is allowed to settle to the $S$ steady state shown in (b). This procedure is repeated once more at $p = 0.225 \approx p_1$, and the $S$ solution initial condition settles to an $H^+$ steady state, shown in (c).}
\label{fig:vonH_bifdiag1}\end{figure}

By modifying the value of $\delta$ in the default parameter set, we find that \eqref{eq:vonH} is amenable to weakly nonlinear analysis in a neighborhood of $\delta = \delta_c \approx 24.7$ (the two Turing points collapse to a single degenerate point at $\delta = \delta_c$). This is observed through the coefficients $b$ and $c$ of \eqref{eq:bifeqns1}. We saw in case (i) of Section \ref{sec:degeneracy} that both small-amplitude stripes and hexagons can be stable when $0<b<c$. Plotting $b$ and $c-b$ evaluated at Turing points $(p_1,q_1)$ and $(p_2, q_2)$ as functions of $\delta$ in Figure \ref{fig:vonH_coeffs}, we observe that $0<b<c$ when $\delta \lesssim 30.0$. Plotting $a$ in a similar way shows that $a$ changes sign between the Turing points when $\delta \gtrsim 25.5$, which permits the assymetric sequences described in Section \ref{sec:degeneracy}, i.e. sequences that involve both gap and spot patterns. Together, these coefficients indicate that the sequence ``$H^- \to S \to H^+$'' appears at small amplitude in the interval $\delta \in [25.5,30.0]$, which parallels what is observed at large amplitude for larger values of $\delta$. A bifurcation diagram depicting this scenario is shown in Figure \ref{fig:vonH_bifdiag1}.

To simulate this sequence numerically, we set $\delta = 27.0$ so that two Turing points emerge at $(p_1,q_1) \approx (0.225, 0.214)$ and $(p_2, q_2) \approx (0.292, 0.247)$. A numerical solution of \eqref{eq:vonH} at $p\approx p_2$ is evolved using a pseudospectral RK4 scheme from spatially random initial conditions for $n$ and $w$ drawn uniformly from the interval $[0.2,0.4]$. After this solution reaches a steady state, the parameter $p$ is decreased in two steps, and the solution is allowed to reach a steady state after each step. These steady states, shown in Figure \ref{fig:vonH_bifdiag1}, closely resemble $H^-$, $S$, and $H^+$ solutions at $p \approx p_2, (p_1+p_2)/2$, and $p \approx p_1$ respectively, which is consistent with the picture suggested by analysis.

\begin{figure}\centering
\includegraphics[width=\columnwidth]{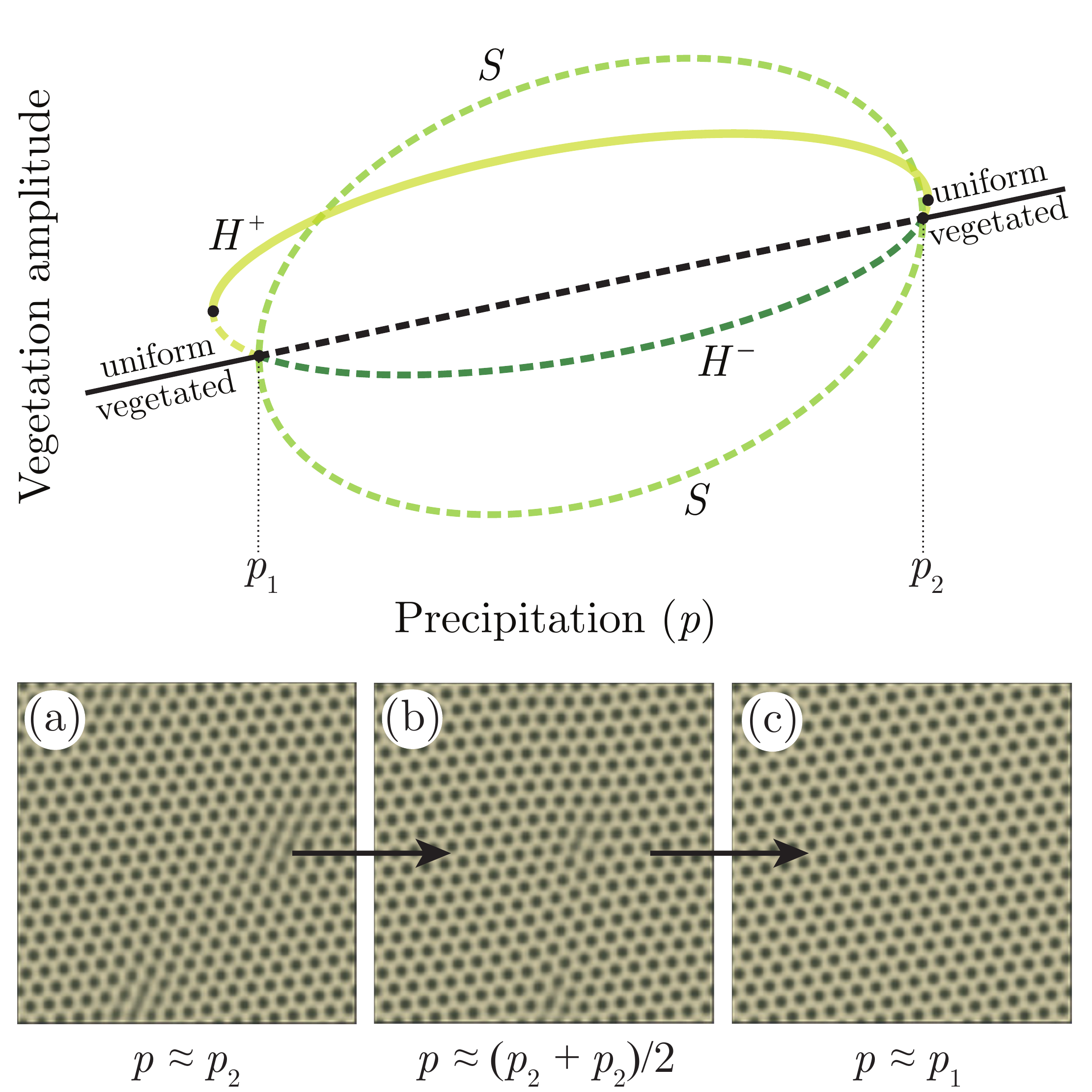}
\caption{(Color online) Bifurcation diagram for solutions of \eqref{eq:vonH} and numerical solutions with $\delta = 25.0$ (default parameter values used otherwise). The numerical simulation (a) is evolved to a steady state resembling an $H^+$ solution from spatially noisy initial conditions at $p = 0.269 \approx p_2$. $p$ is then decreased to $0.257 \approx (p_1 + p_2)/2$ and the $H^+$ state remains stable, as shown in (b). This procedure is repeated once more at $p = 0.245 \approx p_1$, and the $H^+$ state continues to remain stable, as shown in (c).}
\label{fig:vonH_bifdiag2}\end{figure}

Figure \ref{fig:vonH_coeffs} also shows that $a<0$ at both Turing points when $\delta \in [24.7,25.5]$, resulting in ``symmetric'' pattern transition sequences that exclude stable $H^-$ (gaps) patterns in this interval. In Figure \ref{fig:vonH_bifdiag2}, we plot a bifurcation diagram of solutions to \eqref{eq:vonH} as functions of $p$ with $\delta = 25.0$, which shows that only $H^+$ (spot) solutions are stable at small amplitude between the two Turing points. For this value of $\delta$, the two Turing bifurcations occur at $(p_1, q_1) \approx (0.245, 0.231)$ at $(p_2, q_2) \approx (0.269, 0.243)$. Numerical simulations at $p = p_1, (p_1+p_2)/2,$ and $p_2$ all yield small amplitude up-hexagon solutions (see Figure \ref{fig:vonH_bifdiag2}). We remark that although this analysis is evidence of an alternative to the standard sequence in the model by von Hardenberg \textit{et al.}, this alternative occurs only in a small interval of $\delta$.

\section{Discussion}\label{sec:discussion}
In order to analyze aspects of transition between patterned states in PDE vegetation models, we have formulated a bifurcation problem on a 2D hexagonal lattice. Amplitude equations capture dynamics that are dominated by the evolution of critical Fourier mode perturbations to a uniform equilibrium. We enforced degeneracies of the amplitude equation coefficients that enable transitions between patterned states to occur at small amplitude, where they can be investigated analytically. We found that a number of scenarios beyond the standard ``gaps $\to$ labyrinth $\to$ spots'' transition sequence are possible in this generic setting, and that the appearance of these scenarios is distinguished by the coefficients of the amplitude equations near the degenerate point. Since the bifurcation problem exploits symmetries present in a number of vegetation pattern models near Turing bifurcation points, one can place these models within the framework we have described. We did this for the model by von Hardenberg \textit{et al.} \cite{vonHardenberg:2001bka} and observed the standard sequence as well as a ``spots-only'' scenario for two near-degenerate parameter sets. 

Based on our analysis and observations of patterned states in \eqref{eq:vonH}, we speculate that specific assumptions make the standard sequence relatively robust in some PDE models for vegetation pattern formation. This is despite our observation of the ``spots-only'' scenario in \eqref{eq:vonH}, which we found to occur only for a small interval of the parameter we vary. In Section \ref{sec:degeneracy}, we found that an analog of the standard sequence occurs at small amplitude due in part to the quadratic coefficient of \eqref{eq:bifeqns1} changing sign between the Turing bifurcation points. This weakly nonlinear assessment forms an organizing center for more strongly nonlinear model behavior away from a degeneracy, and we conjecture that the quadratic coefficient changing sign between Turing points serves as a crude signpost for the standard sequence. Our expectation is that if a small-amplitude analog of this sequence is observed near a degeneracy, then the quadratic coefficient will often give a qualitative description of behavior even when no small-amplitude patterned states are stable away from the degeneracy. In this way, our approach suggests a framework for assessing the robustness of the standard sequence under parameter variations in \eqref{eq:vonH}, as well as in other models.

Patterned vegetated states often occur during the process of desertification, and the transition between patterned and desert states is accordingly of great interest. In this paper, we presented a picture of pattern formation and collapse that occurs via two Turing bifurcations, but finite-wavelength Turing instability is not the only mechanism for vegetation collapse observed in models. In a study of the generalized Klausmeier model by van der Stelt \textit{et al.} \cite{vanderStelt:2012gsa}, a patterned vegetated state collapses directly to desert via long-wavelength, sideband, and Hopf instabilities. Additionally, homoclinic snaking \cite{Burke:2006kv, Burke:2007kz} has been proposed as a mechanism for the stabilization and motion of localized patterned states that emerge en route to desertification \cite{Bel:2012jqa}. These approaches, as well as the approach we presented in this paper, can help to form a catalog of scenarios for transition between vegetated and desert states in semi-arid ecosystems, which may provide crucial insight into how semi-arid ecosystems will respond to the change in precipitation conditions that will accompany global climate change in the coming decades. 


\section*{Acknowledgements}
This project benefitted from conversations with D. Abrams and E. Meron. Research was supported in part by the NSF Math and Climate Research Network (DMS-0940262).
\bibliography{library}

\end{document}